\documentclass[lettersize,journal]{IEEEtran}
\usepackage{amsmath,amsfonts}
\usepackage{algorithmic}
\usepackage{algorithm}
\usepackage{array}
\usepackage[caption=false,font=normalsize,labelfont=sf,textfont=sf]{subfig}
\usepackage{textcomp}
\usepackage{stfloats}
\usepackage{url}
\usepackage{verbatim}
\usepackage{graphicx}
\usepackage{cite}
\usepackage{amsthm}
\newtheorem*{remark}{Remark}
\def\BibTeX{{\rm B\kern-.05em{\sc i\kern-.025em b}\kern-.08em
    T\kern-.1667em\lower.7ex\hbox{E}\kern-.125emX}}
\begin{document}

\title{A Deep Learning Approach for Pixel-level Material Classification via Hyperspectral Imaging}

\author{Savvas~Sifnaios, George~Arvanitakis, Fotios~K.~Konstantinidis, Georgios~Tsimiklis, Angelos~Amditis, Panayiotis~Frangos
\IEEEcompsocitemizethanks{\IEEEcompsocthanksitem S. Sifnaios, G. Arvanitakis, F. K. Konstantinidis, G. Tsimiklis and A. Amditis are with Institute of Communication and Computer Systems, National Technical University of Athens,9 Iroon. Polytechniou Str., Zografou Athens GR-157 73, Greece.
\IEEEcompsocthanksitem S. Sifnaios and P. Frangos are with School of Electrical and Computing Engineering, National Technical University of Athens,9 Iroon. Polytechniou Str., Zografou Athens GR-157 73, Greece.
\IEEEcompsocthanksitem G. Arvanitakis is with Technology Innovation Institute, Yas Island, Abu Dhabi,United Arab Emirates 
\IEEEcompsocthanksitem S. Sifnaios is the corresponding author (savvas.sifnaios@iccs.gr)}

\thanks{Manuscript received November , 2024; revised  XX, 202X.}
}

\markboth{SUBMITTED IN IEEE TRANSACTIONS ON IMAGE PROCESSING}
{Sifnaios \MakeLowercase{\textit{et al.}}: A Deep Learning Approach for Pixel-level
Material Classification via Hyperspectral Imaging}


\maketitle

\begin{abstract}
Recent advancements in computer vision, particularly in detection, segmentation, and classification, have significantly impacted various domains. However, these advancements are tied to RGB-based systems, which are insufficient for applications in industries like waste sorting, pharmaceuticals, and defense, where advanced object characterization beyond shape or color is necessary. Hyperspectral (HS) imaging, capturing both spectral and spatial information, addresses these limitations and offers advantages over conventional technologies such as X-ray fluorescence and Raman spectroscopy, particularly in terms of speed, cost, and safety.

This study evaluates the potential of combining HS imaging with deep learning for material characterization. The research involves: i) designing an experimental setup with HS camera, conveyor, and controlled lighting; ii) generating a multi-object dataset of various plastics (HDPE, PET, PP, PS) with semi-automated mask generation and Raman spectroscopy-based labeling; and iii) developing a deep learning model trained on HS images for pixel-level material classification. The model achieved 99.94\% classification accuracy, demonstrating robustness in color, size, and shape invariance, and effectively handling material overlap. Limitations, such as challenges with black objects, are also discussed. Extending computer vision beyond RGB to HS imaging proves feasible, overcoming major limitations of traditional methods and showing strong potential for future applications.
\end{abstract}

\begin{IEEEkeywords}
Hyperspectral Imaging, Deep Learning, Material Classification, Pixel-level Classification, Real-time Object Detection
\end{IEEEkeywords}

\section{Introduction}
\label{sec:introduction}
\IEEEPARstart{T}{he} field of material classification has evolved significantly over the past few decades, transitioning from traditional techniques to the application of deep learning methodologies. Traditional methods, such as thresholding, edge detection, and classical machine learning algorithms (e.g., k-nearest neighbors, support vector machines), rely heavily on manually crafted features and heuristic rules. These methods are often effective for simple tasks but struggle with complex and fine-grained material differentiation due to their inherent limitations in capturing shape and colour variations  .

Image classification and segmentation have undergone a paradigm shift with the introduction of deep learning, especially convolutional neural networks (CNNs). Material classification tasks have improved in accuracy and robustness through deep learning models that automatically learn hierarchical features from raw data. However, despite these advancements, challenges remain when using conventional RGB images, especially in scenarios were the color or the shape of the object is not that informative regarding its material.

\begin{figure}[!ht]
    \centering
    \includegraphics[width=\columnwidth]{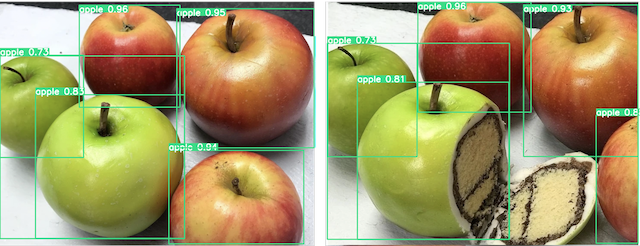}
    \caption{Example of RGB-based object misclassification. This image depicts  pastry cakes that reassemble apples leading the RGB-based model to mistakenly classify the samples as apples.}
    \label{fig:rgb-fail}
\end{figure}

RGB images solely rely on spatial features—such as edges, textures, and colours visible to the human eye. This reliance often leads to misclassification when objects of different classes have similar appearances, as shown in Figure \ref{fig:rgb-fail}. For instance, in the sorting industry, accurately distinguishing between materials with subtle colour differences or complex shapes is crucial for efficiency and quality control. In the pharmaceutical industry, ensuring the purity and correct identification of compounds necessitates fine segmentation capabilities \cite{khan2018modern}. Similarly, in agriculture, RGB imaging often falls short in detecting diseases in crops that present with subtle spectral variations not visible in the RGB spectrum \cite{lu2020recent}. In defense applications, accurate material classification can aid in the detection and identification of hazardous substances, ensuring safety and operational efficiency. 

Hyperspectral imaging has emerged as a promising solution to these challenges. Unlike traditional RGB imaging, hyperspectral cameras capture a wide spectrum of light beyond the visible range, providing detailed spectral information for each pixel in an image. This richness in spectral data allows for more precise material classification and segmentation, overcoming the limitations of traditional methods. 

%The SPECIM FX17 hyperspectral camera, for example, offers high spectral resolution and is capable of capturing up to 400 lines per second, making it suitable for real-time applications  .

Despite its potential and the direction of mobile manufactures' to add more spectrum bands on their phone's cameras \cite{diaz2019real,sharma2023mobispectral,lim2022feasibility}, the research on computer vision beyond RGB imaging is very limited mainly because of the \textbf{lack of publicly available datasets} and the need for \textbf{specialized hardware} that up to this day is expensive. In this paper, we propose P1CH (Pixel-wise 1D Convolutional Hyperspectral) Classifier, a lightweight convolutional neural network designed to perform image segmentation and classification using hyperspectral images. The spectral information captured by the employed hyperspectral camera ranges from 900nm up to 1700nm. The proposed approach leverages the spectral richness of hyperspectral data to enhance the accuracy of material classification, addressing the limitations of traditional methods.
This advancement opens new possibilities for applications in various industries, providing a practical solution to the longstanding problem of precise material classification and segmentation. 
The contributions of this work are summarized as follows:

\begin{itemize}    
    
    \item \textbf{Dataset:} The generation and release of a dataset, containing hyperspectral images and their corresponding labeled masks,  of plastic samples (HDPE, PET, PP, PS), with the aim to avail  research community to investigate other models, applications, insights in the field of computer vision via hyperspectral imaging. 
    
    \item \textbf{P1CH:} The design and implementation of a lightweight yet accurate (99.94\% accuracy) deep learning approach, for pixel-level material classification.

    \item \textbf{Insights:} Extensive performance analysis highlighting the capacity (accuracy, color invariance, shape invariance) as well as the limitations (e.g. black plastics) of the computer vision with hyperspectral imaging.
  
    \item \textbf{Cost:} Taking advantage of the deep learning capabilities, cost-efficient alternatives for calibration and normalization of Hyperspectral images are proposed tailored for deep learning models.

\end{itemize}

To this end, Section~\ref{sec:related work} describes the related work, while Section~\ref{sec: hs imaging setup} provides a high-level description of the system setup and the hardware equipment used in this work. The creation of hyperpsectral dataset, the spectral preprocessing methodologies along with the semi-automated AI-assisted mask generation are presented in Section \ref{sec:dataset}. Section~\ref{sec:model-architecture-training} introduces the architecture of the proposed model, the preparation of the training instances. The training process process, along with the selected hyperparameters and the initial results are also discussed in that section. In addition, the experimental results on various scenarios, i.e. \textit{mixed materials}, \textit{shredded samples} and \textit{overlapping objects}, and the performance of the proposed system are discussed in Section~\ref{sec:results}. This section, also, explores the limitations of the hyperspectral model in the case of dark-coloured or black samples. Finally, in Section~\ref{sec:conclusion}  a discussion about the main conclusions of this study and the future steps are provided.

\section{Related Work}
\label{sec:related work}

Material classification has historically relied on several approaches that cut across many domains, including electro-mechanical and chemical analysis techniques. Raman and Near-Infrared spectroscopy, coupled with a multivariate analysis, have widely been used to identify the materials' composition \cite{przestrzelski2018integrating,jang2020korean,penumuru2020identification}. In another work, the utilisation of X-Ray Diffraction (XRD),  Energy Dispersive X-ray Spectroscopy (EDS) and Atomic Absorption Spectroscopy (AAS) techinques have been proposed, because their high sensitivity and accuracy in detecting microstructural features and hence identifying the element composition of a sample \cite{chowdhury2016image}. Zhang et al. \cite{zhang2022machine} highlighted the use of Differential Scanning Calorimetry (DSC) and Thermogravimetric Analysis (TGA) in accurately determining the thermal properties of various materials . Zhang and Shao \cite{zhang2022imagebased} emphasized the role of optical microscopy in material.

In an attempt to further increase the performance and robustness of material classification pipelines, various studies explored the potential of deep learning. In \cite{weiss2019radar} a deep convolutional neural network (CNN) was introduced for classifying 60-GHz radar data with an accuracy of 97\%. Deep neural networks have, also, been utilised to analyse surface haptic data \cite{zheng2016haptic}, achieving a precision of 94\% and recall of 92\%. Moreover, CNNs have been proposed for the case of visual RGB images \cite{zoumpoulis2024smart,zoumpoulis2024advancing,adedeji2019waste,azimi2018steel} in various use cases, e.g. commercial waste, steel products, etc. In another work,  Konstantinids et al. \cite{KONSTANTINIDIS2023107244} proposed a multi-modal deep classifier based on ResNet-18 that jointly extracts information from RGB and multispectral cameras to classify plastic polymers, as well as wood by products with an object accuracy of 96\%.

%Deep learning revolutionized the area of material classification. Different deep learning models have been used to enhance the accuracy and effectiveness of various classification tasks. As an example, Weiß and Santra \cite{weiss2019radar} employed a deep convolutional neural network (CNN) in classifying materials based on 60-GHz radar data with an accuracy of 97\%.Swanson et al. \cite{swanson2020amorphous} developed an amorphous material classifier using deep learning, and reported a precision of 94\% and recall of 92\%. Zheng et al. \cite{zheng2016haptic} employed deep learning to perform surface material classification through haptic and visual information fusion with an accuracy of 96\%. Adedeji and Wang \cite{adedeji2019waste} proposed a Deep CNN-based intelligent waste classification system with an accuracy rate of 98\%. Azimi et al. \cite{azimi2018steel} applied deep learning to classify microstructures in steel materials, achieving an F1-score of 93\%. In another work,  Konstantinids et al. \cite{KONSTANTINIDIS2023107244} proposed a multi-modal deer classifier based on ResNet-18 that jointly extracted information from RGB and multispectral cameras to classify plastic polymers, as well as wood by products with an accuracy of 96\%.

While the cited work above demonstrates impressive classification performance, these methods are often slow and computationally intensive. Hyperspectral imaging has become a powerful means for in situ material classification, since they offer the advantage of real-time and efficient analysis. Shaikh and Thörnberg \cite{shaikh2022impact} investigated the impact of water vapor on polymer identification by means of short-wave infrared hyperspectral imaging yielding an accuracy rate around 88\%. According to Shaban \cite{shaban2013determination}, hyperspectral imaging has been used to determine different characteristics pertaining to concrete without disturbing its structure at sites, where it would not be possible to take samples back to laboratories for analysis; the typical accuracies are above 90\%. With about 93\% accuracy Capobianco et al.\cite{capobianco2015hyperspectral} have characterized ancient roman wall paintings using Hyperspectral Imaging as well as aid in artwork authentication (Polak et al. \cite{polak2017authentication}), achieving a precision of 95\%.

The combination of deep learning and hyperspectral imaging has shown great promise in improving the material classification capability in Earth Observation applications. Notable examples include models, e.g. Xception-based, CNN-based systems, and R-VCANet achieving accuracies up to 99\%, with precision and recall values around 94\% and 93\%, and F1 score reaching 91\% \cite{banumathi2021intelligent, li2019deep, windrim2018pretraining, chen2019automatic, pan2017rvcnet}. Venkatesan et al. \cite{venkatesan2019hyperspectral} applied deep recurrent neural networks in medical hyperspectral images to achieve feature recognition with a precision rate of about 96\%. Xiong et al. \cite{xiong2020tracking} developed material tracking methods for hyperspectral videos using deep learning, achieving an F1 score of 94\%. Medus et al. \cite{medus2021cnn} applied CNNs to classify hyperspectral images in industrial food packaging, reporting an accuracy of 99\%. Okada et al. followed a patch-wise approach for the identification of 5 different mineral types, using a VGG16-based CNN to classify the acquired HS images they achieved high accuracy over 90\% \cite{min10090809}. Extreme Learning combined with Stacked Autoencoders for feature extraction has also been applied on HS images, in \cite{9094372}, for the detection of plastic films within cotton feed stock, creating a pixel-level classification map with accuracy up to 95\%. Zhu et al. in \cite{molecules24183268}, also attempted to classify different cotton seeds. Their work encompasses a multivariate analysis for manual extraction of the 10 most informative features, which were subsequently fed to CNN for the classification task, resulting in a prediction accuracy of 88\%. Moreover, Artzai et al. worked with HS images with 76 bands of non ferrous metals, and utilised them to train a CNN-based U-Net like network \cite{picon2024hyperspectraldatasetdeeplearning}. This approach analysed the HS images as a whole and created a classification map with an accuracy equal to 95\%. Finally, Roy et al.\cite{roy2019hybridsn} in their work proposed a methodology that incorporates PCA as a preprocessing step with a deep convolutional network, i.e. HybridSN, for the pixel-wise classification of Hyperspectral Images, achieving an accuracy score of 99\%.

The aforementioned studies with hyperspectral classification using deep learning have yielded remarkable results. However, it is important to note that most of the already limited approaches that exist in the field commonly either deal with HS images in batches or utilise 2D Convolutional Neural Networks (CNNs). In order to achieve this, they use 2D convolutional kernels that inevitably lead to loss of accuracy at object borders. Notably, for 2D CNNs convolutions are designed to capture spatial context within each individual frame but this may blur boundaries or misclassify pixels on the boundary between neighboring objects that have different spectral signatures. This means that spectral continuity across edges is not incorporated by the 2D convolution operation, causing edge artifacts and reducing classification accuracy here. These limitations highlight the need for more refined techniques that analyse HS images on a pixel level and can preserve boundary integrity while leveraging the rich spectral data provided by hyperspectral imaging.

%Banumathi and Muthumari \cite{banumathi2021intelligent} developed a deep learning-based Xception model aimed at analyzing and classifying hyperspectral images, with an accuracy of 98\%. Moreover, Li et al. \cite{li2019deep} gave a summary of advanced techniques for HSI classification using deep learning methods which led to accuracy ranging from 85\% to 99\%  according to the dataset used as well as the model. The authors in Windrim et al. \cite{windrim2018pretraining} investigated benefits pre-training on CNN-based systems for HS image classification giving out precision and recall values at 94\% and 93\% consecutively. Additionally, Chen et al. \cite{chen2019automatic} proposed an automated configuration of CNNs that achieved an F1-score value of 91\%. Pan et al. \cite{pan2017rvcnet} introduced R-VCANet which is a new deep learning based method for HSI classification with an accuracy of about 97\%. 

\section{Hyperspectral Imaging Setup}
\label{sec: hs imaging setup}
This section provides a detailed description of the physical infrastructure of the cyber-physical system where the dataset was created using materials that were conveyed under the vision and hyperspectral sensors. 

\subsection{Imaging Spectroscopy}
Hyperspectral imaging combines conventional imaging and spectroscopic methodologies with the goal to simultaneously obtain spatial and spectral information from various wavelengths of the electromagnetic spectrum for every individual pixel in an image of a scene, with the objective of locating objects, classifying materials, or detecting processes~\cite{brady2009optical}. Pushbroom sensors capture spectral information across a swath as the sensor moves, line scan sensors capture data one line at a time, while whiskbroom sensors, scan point-by-point to build an image, and snapshot sensors capture the entire spectral image in a single exposure.

\subsection{Camera}
\label{subsec:camera}

The SPECIM FX17 line scan camera, following push broom technology, utilises a matrix detector and an imaging spectrograph to capture spectral data efficiently. Light enters through high-performance optics and an entrance slit, forming a line image that the spectrograph disperses into a spectrum (900 - 1700 nm, across 224 bands). This setup allows each axis of the detector to record spatial position and spectral information simultaneously. It ensures measurement stability despite sample or camera movement, requires less illumination power while achieving higher intensity, and is significantly more efficient than filter-based cameras, yielding a purer spectrum.

\subsubsection{Conveyor}
Once the line scan camera is selected for capturing, the system comprises a conveyor belt that facilitates the horizontal movement of objects at an adjustable speed, in order to achieve synchronised sampling frequency of the hyperspectal sensor and objects' movements. The camera was placed 0.73 cm from the conveyor, while the illumination source was placed 0.5 meters vertically above the conveyor belt, at a 45-degree angle, and 0.3 meters horizontally from the center of the camera.

\subsection{Illumination}

The delpoyed custom-made LDL-222X42CIR Full-Spectrum bar light manufactured by CCS Inc., offers better performance for Imaging spectroscopy challenges in contrast with Led lights~\cite{sifnaios2023exploration}. Specifically, it includes four different halogen bulbs that emit light in distinct regions of the spectrum and a power rating of 87W. This permits it to span the complete range of wavelengths from 400nm to 2400nm. This light source contains a dispersion layer that uniformly distributes the output light in multiple directions, resulting in more consistent illumination of the scene and the objects within it.

\subsection{Acquired Data}

The HS camera employed in this work is ample of achieving a maximum sampling rate of 400 FPS, with each scanned line being of shape \(640\times224\). Moreover, camera provides the option of applying on-chip spatial and/or spectral binning, hence reducing the respective dimensions by factors of \(\times2\), \(\times4\), or \(\times8\). In this work, no spectral or spatial binning was applied, thus the acquired data were of shape \(n_{rows}\times640\times224\), where \(n_{rows}\) is depended on the selected sampling rate and the acquisition duration. Finally, it is worth mentioning that the pixel size of the selected camera is 0.9375mm, thus allowing for very precise and accurate segmentation. 

\section{Dataset}
\label{sec:dataset}
\begin{figure}[!ht]
    \centering
    \includegraphics[width=0.42\textwidth]{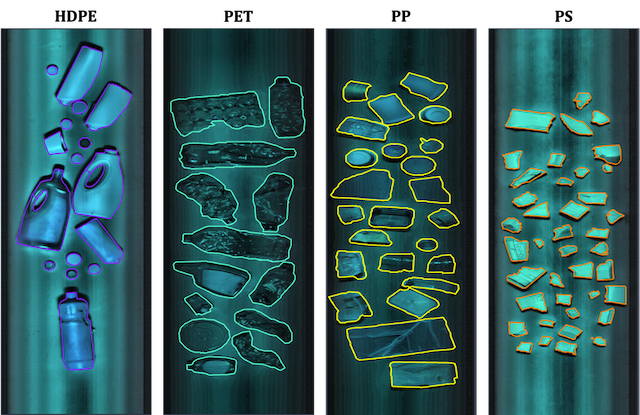}
    \caption{The false colour, contrast stretched HS images used for the generation of the training set. }
    \label{fig:trainig-images}
\end{figure}
Following the aforementioned hyperspectral imaging setup, this section presents the dataset used in the experimental part of this work. At first, a comprehensive description of the dataset is provided, highlighting the different material classes involved in the dataset and summarizing the number of images, objects, and pixels for each category. The following subsections address the processes involved in dataset preparation, including the normalization of the images to ensure numerical consistency. The generation of ground truth masks is subsequently discussed, pointing out their necessity for the proposed supervised learning approach. Moreover, the reasoning behind the utilization of Raman spectroscopy throughout labeling process is explained. Finally, the generation of the training batches and their format is outlined.

\subsection{Dataset description}

For the generation of the train and test dataset an ensemble of plastic samples of different material categories were collected consisting of \textit{HDPE}, \textit{PET}, \textit{PP}, and \textit{PS} objects.

\textbf{Train Set:} In detail, \textit{169} objects were selected to represent the classes. The physical objects were separated according to their material classes and subsequently placed randomly on the conveyor belt, described in Section \ref{subsec:camera}. For the augmentation of the training set by every sample was  captured twice with slightly different light conditions. A comprehensive analysis of the dataset composition is presented in Table \ref{tab:summary}. 
%As mentioned above, 2 images are acquired per plastic's category, of size \(n_{rows}\times640\times224\), where \(n_{rows}\) is variable depending of the length of the scene, \(640\) is each image's width, while \(224\) are the respective spectral channels. 

%Within this images \textit{80 HDPE}, \textit{47 PET}, \textit{89 PP} and \textit{120 PS} samples are depicted. On first sight, the dataset might appear as imbalanced, however, since the proposed model is a pixel-wise classifier what is crucial for this application is the number of pixels per class. To this end, utilizing the ground truth masks, described later in subsection \ref{subsec:masks}, the number of pixels per class is calculated. Specifically, \textit{HDPE} is depicted in \textit{794,142} pixels, \textit{PET} in \textit{810,483} pixels, \textit{PP} in \textit{831,340} pixels, while \textit{PS} is depicted in \textit{801,936} pixels, hence leading to the conclusion that the required class balance is achieved on the pixel-level. 
  
\begin{table}[ht]
    \centering
    \caption{Summary of Images, Objects, and Pixels per Category of the Training set.}
    %\resizebox{0.9\columnwidth}{!}{%
        \begin{tabular}{|c|c|c|c|}
            \hline
            \textbf{Material} & \textbf{ Images} & \textbf{ Objects} & \textbf{Pixels} \\ \hline
            HDPE & 2 & 40 & 794,142 \\ \hline
            PET & 2 & 24 & 810,483 \\ \hline
            PP & 2 & 45 & 831,340 \\ \hline
            PS & 2 & 60 & 801,936 \\ \hline
        \end{tabular}
%    }
    
    \label{tab:summary}
\end{table}

At this point, it is important to mention that a \(5^{th}\) class was introduced to the dataset. Namely, this class is \textit{Background} and it represents the set of pixels depicting the conveyor belt's surface. Since the conveyor belt is apparent in every HS image, no matter the type of plastic, initially the number of \textit{Background} pixels was \(\times10\) higher than any other class, potentially leading the model to be biased in favor of this class. To handle this, a random subset of \textit{Background} pixels was selected, ensuring its size will be equal to the maximum number of pixels of the plastic classes, i.e. 831,340 pixels.

\textbf{Test Set:} Regarding the test set, was prepared from different objects than the train set, but that belong to the same material categories. Additionally the dataset was selected in order to capture four main cases i) \textit{unmodified objects}, similar to the train set ii) \textit{shredded objects}, the shape do not capture any information iii) Mixed-overlapping materials, where a material is into another material, Table \ref{tab:without_black_pixels} presents the dimensions of the test set.

% \begin{table}[ht]
%     \centering
%     \caption{Material classification without black pixels}
%     \begin{tabular}{|c|c|c|}
%         \hline
%         \textbf{Material} & \textbf{Objects} & \textbf{Pixel} \\ \hline
%         HDPE & 29 & 124,109 \\ \hline
%         PET  & 20 & 100,445\\ \hline
%         PP  & 2 & 59,682 \\ \hline
%         PS  & 29 & 116,704 \\ \hline
%     \end{tabular}

%     \label{tab:without_black_pixels}
% \end{table}

\begin{table}[ht]
    \centering
    \begin{minipage}{0.45\columnwidth}
        \centering
        \caption{Test set without black samples.}
        \begin{tabular}{|c|c|c|}
            \hline
            \textbf{Material} & \textbf{Objects} & \textbf{Pixel} \\ \hline
            HDPE & 29 & 124,109 \\ \hline
            PET  & 20 & 100,445 \\ \hline
            PP   & 2  & 59,682  \\ \hline
            PS   & 29 & 116,704 \\ \hline
        \end{tabular}
        \label{tab:without_black_pixels}
    \end{minipage}
    \hspace{0.05\columnwidth}
    \begin{minipage}{0.45\columnwidth}
        \centering
        \caption{Black samples test   \\subset.}
        \begin{tabular}{|c|c|c|}
            \hline
            \textbf{Material} & \textbf{Objects} & \textbf{Pixels} \\ \hline
            HDPE & 19 & 31,649 \\ \hline
            PET  & 0  & 0 \\ \hline
            PP   & 14 & 25,280 \\ \hline
            PS   & 5  & 5,035 \\ \hline
        \end{tabular}
        \label{tab:only_black_samples}
    \end{minipage}
\end{table}

Furthermore, regarding the literature \cite{blackPlastics,Lindell2018}, HS cameras (up to 1700nm) cannot capture information regarding plastics that are coloured black. In order to examine that an extra black plastic dataset was collected and presented in the Table \ref{tab:only_black_samples}

% \begin{table}[ht]
%     \caption{black samples test dataset}
%     \centering
%     \begin{tabular}{|c|c|c|}
%         \hline
%         \textbf{Material} & \textbf{Objects} & \textbf{Pixels} \\ \hline
%         HDPE & 19 & 31,649 \\ \hline
%         PET & 0 & 0 \\ \hline
%         PP  & 14 & 25,280 \\ \hline
%         PS  & 5 & 5,035\\ \hline
%     \end{tabular}
%     \label{tab:only_black_samples}
% \end{table}

\begin{figure*}[!ht]
    \centering
    \includegraphics[width=1\textwidth]{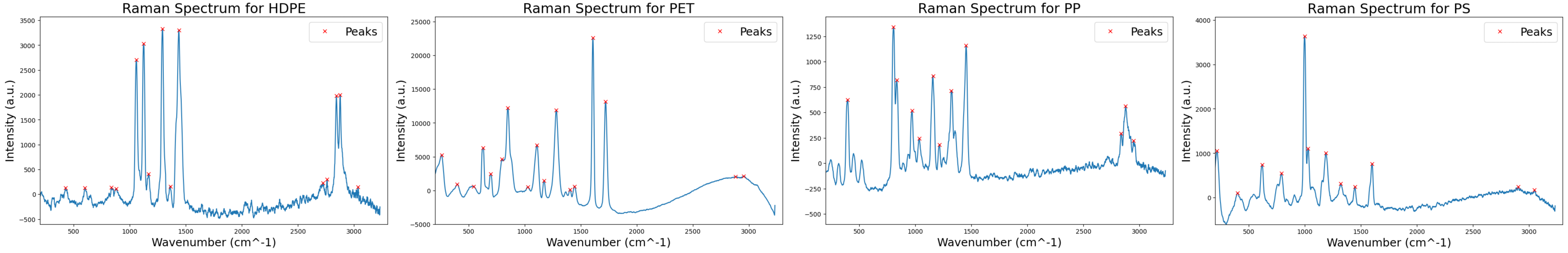}
    \caption{The Raman spectra for the 4 material classes. For the HDPE spectrum, the peaks at 1058 cm\textsuperscript{-1}, 1123 cm\textsuperscript{-1}, 1291 cm\textsuperscript{-1}, 1437 cm\textsuperscript{-1}, and the range from 2843-2876 cm\textsuperscript{-1} are characteristic of high-density polyethylene (HDPE). These peaks correspond to the vibrations of the molecular structure of HDPE, specifically indicating the various stretching and bending vibrations of the C-H bonds. Accordingly, for the rest of the spectra, it is pointed out that two characteristic peaks at 1607 cm\textsuperscript{-1} and 1721 cm\textsuperscript{-1} were detected in the top-right plot and correspond to the vibrations of the phenyl group in the polyester. Additionally, the peaks in the range of 1100-1200 cm\textsuperscript{-1} indicate the stretching vibrations of the C-O group. The peaks observed at 970 cm\textsuperscript{-1}, 1034 cm\textsuperscript{-1}, 1360 cm\textsuperscript{-1}, 1453 cm\textsuperscript{-1}, and 2946 cm\textsuperscript{-1} in the bottom-left plot are associated with the vibrations of the methyl group (CH\textsubscript{3}) in polypropylene, while the intense peak observed at 1010 cm\textsuperscript{-1}, in the bottom-right plot, along with the peak at 1598 cm\textsuperscript{-1}, suggest the presence of polystyrene.}
    \label{fig:raman}
\end{figure*}
\subsection{Acquisition Pipeline}

This section describes the process of converting the captured pixel values from \textit{16-bit} \textbf{unsigned} integers, in the range of \([0,4095]\) to compatible for deep learning and ready-to-used \textit{32-bit} floats pixel values in the range of \([0.,1.]\).

Throughout the literature it is strongly suggested a) to use the \textbf{spectral calibration matrix}, that is specifically provided by the camera manufacturer, and post the HS image's multiplication with spectral calibration matrix, b) to \textbf{normalize} the calibrated data using \textbf{White} and \textbf{Black} reference. 

%In this subsection, the data acquisition and preparation pipeline is described in depth. Up to this point, raw Hyperspectral data have been collected from the camera's sensor. In order for the data to become compatible with the data requirements set by the neural architecture, a normalization algorithm needs to be utilised to transform the data from \textit{unsigned 16-bit integers} to \textit{32-bit floats} in the range of \([0,1]\).

%Throughout the literature for processing HS images, the normalization of each image using a black and a white reference is underlined as a crucial step in the pipeline. In order to apply the black-white reference normalization, a white reflection target is required. Such equipment is often not commercially available and its cost grows exponentially with respect to its dimensions. 

\textbf{White reference:} A white reflection target is required with the key property of reflecting the incident radiation uniformly across all wavelengths. In this manner, the real maximum absolute pixel value a HS camera can capture, given the illumination source, is calculated and ultimately creates the white reference. Although should be taken into account that a white reflection target is often not commercially available and its cost grows exponentially with respect to its dimensions.

\textbf{Black reference:} On the other hand, the black reference is utilised to model the sensors' electronic noise, caused by the electrons' random movement due to the sensors' temperature. In this work, the black reference \(I_{\text{black}}(\lambda)\) was acquired, by closing the shutter of the camera and capturing 1000 lines, which were then averaged to create the black reference. For the maximum absolute pixel value the HS camera can capture, the assumption was made that it can be approximated by calculating the maximum pixel value within the train set.

By modeling that noise, one can subtract the black reference both from the HS image and the white reference, to acquire the noise-corrected version of both images. Equation \ref{eq:white-black} provides a mathematical formulation of the black-white reference normalization: 

\begin{equation}
    \label{eq:white-black}
    I_{\text{norm}}(x, y, \lambda) = \frac{I(x, y, \lambda) - I_{\text{black}}(\lambda)}{I_{\text{white}}(\lambda) - I_{\text{black}}(\lambda)} ~~, 
\end{equation}

where \(I_{\text{norm}}(x, y, \lambda)\) is the normalized hyperspectral image at pixel \((x, y)\) and wavelength \(\lambda\), \(I(x, y, \lambda)\) is the raw hyperspectral image at pixel \((x, y)\) and wavelength \(\lambda\),
\(I_{\text{black}}(\lambda)\) is the black reference image at wavelength \(\lambda\),
\(I_{\text{white}}(\lambda)\) is the white reference image at wavelength \(\lambda\).

The black reference \(I_{\text{black}}(\lambda)\) was acquired, by closing the shutter of the camera and capturing 1000 lines, which were then averaged to create the black reference. For the maximum absolute pixel value the HS camera can capture, the assumption was made that it can be approximated by calculating the maximum pixel value within the train set. Equation \ref{eq:our-black-white} provides an approximation of Equation \ref{eq:white-black} that does not require an expensive white reflection target:

\begin{equation}
    \label{eq:our-black-white}
    I_{\text{norm}}(x, y, \lambda) = \frac{I(x, y, \lambda) - I_{\text{black}}(\lambda)}{\text{M}} ~~,
\end{equation}

where  \(I_{\text{norm}}(x, y, \lambda)\) is the normalized hyperspectral image at pixel \((x, y)\) and wavelength \(\lambda\),
\(I(x, y, \lambda)\) is the raw hyperspectral image at pixel \((x, y)\) and wavelength \(\lambda\),
\(I_{\text{black}}(\lambda)\) is the black reference image at wavelength \(\lambda\),
\(M\) is the dataset's maximum pixel value.

It should be noted that in HS image processing applications it is highly recommended to use the aforementioned steps for \textit{spectral calibration} and \textit{normalization} which often include expensive equipment and dependencies on the hardware manufacturer. However, as shown in Section \ref{sec:results}, even if we totally ignore those steps, the learning performance of the proposed deep learning algorithm is not affected.

\begin{remark}[Calibration on Training]
The spectral calibration and the normalization as operations are a sequence of matrix multiplications, it seems by our experiments, that this transformation can be learned directly in the training process.
\end{remark}

\subsection{Ground truth mask generation}
\label{subsec:masks}

As in every supervised learning application, a set of ground truth labels is needed to ensure successful training of the neural network model. To this end, an AI-assisted methodology was deployed for the generation of the binary masks, presented in Figure \ref{fig:trainig-images}, which later on will be utilised as training labels. 

\subsubsection*{\textbf{Semi-Automated Segmentation}}
The first step in the proposed methodology, is the creation of a false-colour RGB version of the HS image. To this end, the Standard Deviation of each channel in the original image was calculated, as a measure of its contrast. The three channels with the highest contrast were selected and sorted in ascending order of wavelength, for each image, in order to create the false-colour RGB image. An adaptive histogram stretching algorithm was, also,  applied to the respective RGB versions in order to further increase the contrast and make the objects' edges as sharp as possible without altering the spatial content of each image. 

The false-colour, histogram stretched images were subsequently utilised for the generation of segmentation masks. A ViT model, namely SAM (Segment Anything) \cite{kirillov2023segment}, for the semi-automated mask generation task. In detail, positive (pixel to be included in the mask) and negative  (pixel to be excluded from the mask) points were given as prompts to the model in order to generate a first estimation of the mask. The predicted mask was then visually inspected and refined, when needed, aiming for precision maximization at the boundaries of the object. This procedure was repeated for every image and every object depicted within an image of the dataset and the final results can be seen in Figure \ref{fig:trainig-images}. 

\begin{figure*}[!ht]
    \centering
    \includegraphics[width=0.95\textwidth]{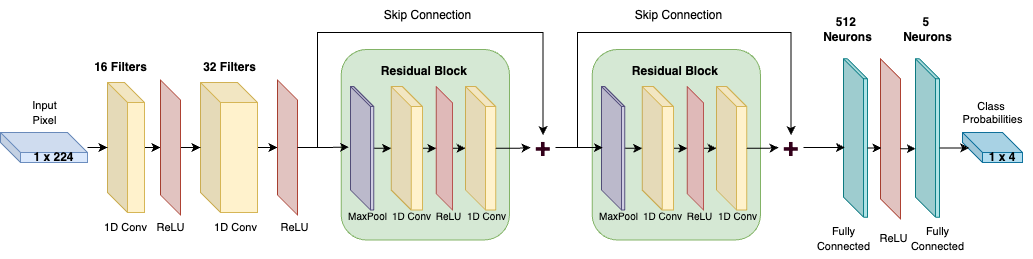}
    \caption{The high-level architecture of the proposed Pixel-wise 1D Convolutional Hyperspectral (P1CH) Classifier .}
    \label{fig:model architecture}
\end{figure*}
\subsubsection*{\textbf{Labeling - Raman}}

The final step of the ground truth generation is to assign a class to each of the aforementioned masks. To this end, Raman Spectroscopy was employed. 

Raman spectroscopy is a powerful analytical technique used to observe vibrational, rotational, and other low-frequency modes in a system. It relies on inelastic scattering, or Raman scattering, of monochromatic light, typically from a laser. When light interacts with molecules, vibrations, or other excitations in the system occur shifting up or down the energy of the laser, creating peaks in the acquired spectrum and hence providing a fingerprint by which molecules can be identified \cite{smith2019modern, long2002raman}. In Figure \ref{fig:raman}, an example of the spectrum for each of the 4 plastic types is presented, where the red marked peaks indicate the existence of each polymer in under examination the plastic sample.

% For the HDPE spectrum, the peaks at 1058 cm\textsuperscript{-1}, 1123 cm\textsuperscript{-1}, 1291 cm\textsuperscript{-1}, 1437 cm\textsuperscript{-1}, and the range from 2843-2876 cm\textsuperscript{-1} are characteristic of high-density polyethylene (HDPE). These peaks correspond to the vibrations of the molecular structure of HDPE, specifically indicating the various stretching and bending vibrations of the C-H bonds. Accordingly, for the rest of the spectra, it is pointed out that two characteristic peaks at 1607 cm\textsuperscript{-1} and 1721 cm\textsuperscript{-1} were detected in the top-right plot and correspond to the vibrations of the phenyl group in the polyester. Additionally, the peaks in the range of 1100-1200 cm\textsuperscript{-1} indicate the stretching vibrations of the C-O group. The peaks observed at 970 cm\textsuperscript{-1}, 1034 cm\textsuperscript{-1}, 1360 cm\textsuperscript{-1}, 1453 cm\textsuperscript{-1}, and 2946 cm\textsuperscript{-1} in the bottom-left plot are associated with the vibrations of the methyl group (CH\textsubscript{3}) in polypropylene, while the intense peak observed at 1010 cm\textsuperscript{-1}, in the bottom-right plot, along with the peak at 1598 cm\textsuperscript{-1}, suggest the presence of polystyrene, as these peaks are attributed to the vibrations of the phenyl group.

In this manner, each sample annotated in the previous step was individually scanned with the Raman equipment and it spectrum was analysed in order to identify the indicative, for each class, peaks. The results of the Raman Spectroscopy analysis, were used as the class of each of the aforementioned masks.

\section{Model Architecture \& Training}
\label{sec:model-architecture-training}
\subsection{Architecture}
\label{subsec:arch}

The proposed hyperspectral image classification model uses a \textbf{1D CNN architecture} to capture and process spectral information. The architecture uses a) 2 convolutional layers, b) 2 residual blocks, and c) 2 fully connected layers to accurately classify hyperspectral data. The architecture of the proposed Pixel-wise 1D Convolutional Hyperspectral Classifier is shown in Figure \ref{fig:model architecture}.

\textbf{Convolutions:} The input pixel of shape $(1 \times 224)$ is passed through an initial convolutional layer with 16 size-3 filters and padding to preserve dimensions. Afterwards, a 32-filter, 3-size convolutional layer with padding follows. After each convolutional layer, a ReLU activation function introduces non-linearity and a max-pooling layer with a kernel size of 2 and stride of 2 reduces data dimensionality.

\textbf{Residuals:} The model uses two residual blocks for feature extraction and learning. The first residual block receives the output from the second convolutional layer and processes it through two convolutional layers with 64 filters each, maintaining a kernel size of 3. Batch normalization is also applied after each convolutional layer. This output is added to the block's input - through the utilisation of skip connection - and passed though a ReLU activation function. The second residual block follows a similar structure, but with 128 filters in each convolutional layer.

\textbf{Fully Connected}: The processed features are flattened and passed through a fully connected layer with 512 neurons, followed by a dropout layer. A second fully connected layer follows with its number of neurons set equal to the number of classes. Finally a soft-max layer provides the final classification probabilities.

\subsection{Data Pre-Processing}

As mentioned in the \textit{Architecture}, the model expects a 1-dimensional vector as input. In this work, the input vectors are the individual pixels of the dataset's images. Up to this point, however, the dataset consists of HS images, hence it is necessary to convert the images into a set of pixels. 

To this end, a data handling pipeline was implemented to efficiently utilise the large volume of data encoded within HS images. In detail, a memory-mapped array uses the operating system's virtual memory capabilities to map a disk file directly into the address space of the application, allowing for efficient, random access to large datasets without loading the entire file into memory. In this manner, by employing memory-mapped arrays out-of-core processing is achieved, which significantly reduces the memory footprint and improves the performance of data loading operations. 

By exploiting the capabilities of memory-mapped arrays, the entire set of images, along with their respective ground truth masks are flattened across the spatial dimensions, thus creating the desired 1-dimensional vectors is 224 features each. The feature vectors are, subsequently, randomly shuffled and split in two subsets; the train and the validation set with ratios 90\% and 10\% respectively. 

\subsection{Model Training}

With the data preprocessing pipeline established, the focus can now shift on the model training phase, where the calibrated, and pre-processed data is being used to train the Pixel-wise 1D Convolutional Hyperspectral Classifier. 

A dataloader was defined for each of the two sub-sets, each with a \textit{batch size} equal to 640. The samples in the train set are shuffled in the beginning of each epoch, thus ensuring slightly different data distribution on batch-level in every iteration. The selected optimizer is Adam \cite{kingma2014adam}, and the initial learning rate was set to \(0.001\). To prevent the model being stuck to a local minima of the loss function, a learning rate (LR) scheduler was also implemented. The LR scheduler utilised in this work was \textit{Cosine Annealing with Warmup} \cite{loshchilov2016sgdr,loshchilov2017decoupled}. The mathematical formulation of this scheduler is described in the following equation : 

\vskip 0.1in
\scalebox{0.92}{
$
\eta_t =
\begin{cases}
    \eta_{\text{init}} \frac{t}{T_{\text{w}}} & \text{if } t \leq T_{\text{w}} \\
    \eta_{\text{min}} + \frac{1}{2} (\eta_{\text{max}} - \eta_{\text{min}}) 
    \left( 1 + \cos\left( \frac{(t - T_{\text{w}}) \pi}{T - T_{\text{w}}} \right) \right) & \text{if } t > T_{\text{w}}~~,
\end{cases}
$
}
\vskip 0.1in
where \(\eta_t\) is the learning rate at epoch \(t\).
\(\eta_{\text{init}}\) is the initial learning rate,
\(\eta_{\text{max}}\) is the maximum learning rate,
\(\eta_{\text{min}}\) is the minimum learning rate,
\(T_{\text{w}}\) is the number of warmup epochs,
\(T\) is the total number of epochs.

Since the model's aim is to classify each of the pixel in a HS image to their respective class, the \textit{Cross-Entropy} Loss Function \cite{goodfellow2016deep} was selected to be minimized throughout the training process as depicted in Equation \ref{eq:cross-entropy}.

%Cross-entropy loss is the preferred choice for classification tasks as it quantifies the dissimilarity between the predicted probability distribution and the true distribution. This loss function is advantageous in deep learning models because it directly optimizes the likelihood of the correct class, thereby facilitating faster convergence and more accurate probability estimation by minimizing the Kullback-Leibler divergence between the distributions \cite{kullback1951information}.  provides a systematic overview of the aforementioned loss function. 

\begin{equation}
\label{eq:cross-entropy}
    \mathcal{L}_{\text{CE}} = -\sum_{i=1}^{N} \sum_{c=1}^{C} y_{ic} \log(\hat{y}_{ic}) ~~,
\end{equation}
where \(N\) is the number of samples,
\(C\) is the number of classes.
\(y_{ic}\) is the ground truth label (1 if sample \(i\) belongs to class \(c\), otherwise 0),
\(\hat{y}_{ic}\) is the predicted probability that sample \(i\) belongs to class \(c\).

The number of training epochs \(T\) was set to 50, while a callback was also employed, in order to save to the model's checkpoint with the best accuracy score in the validation set, thus ensuring that after the training process is finished the final model will be the one with the best performance in the validation set. Moreover, it is noted that the initial and maximum learning rate \(\eta_{\text{init}}\) , \(\eta_{\text{max}}\) respectively were set to \(0.001\), \(\eta_{\text{min}}\) was set to \(0.0001\), while warmup \(T_{\text{w}}\)  was set to 10 epochs. 

\begin{figure}[!ht]
    \centering
    \includegraphics[width=0.95\columnwidth]{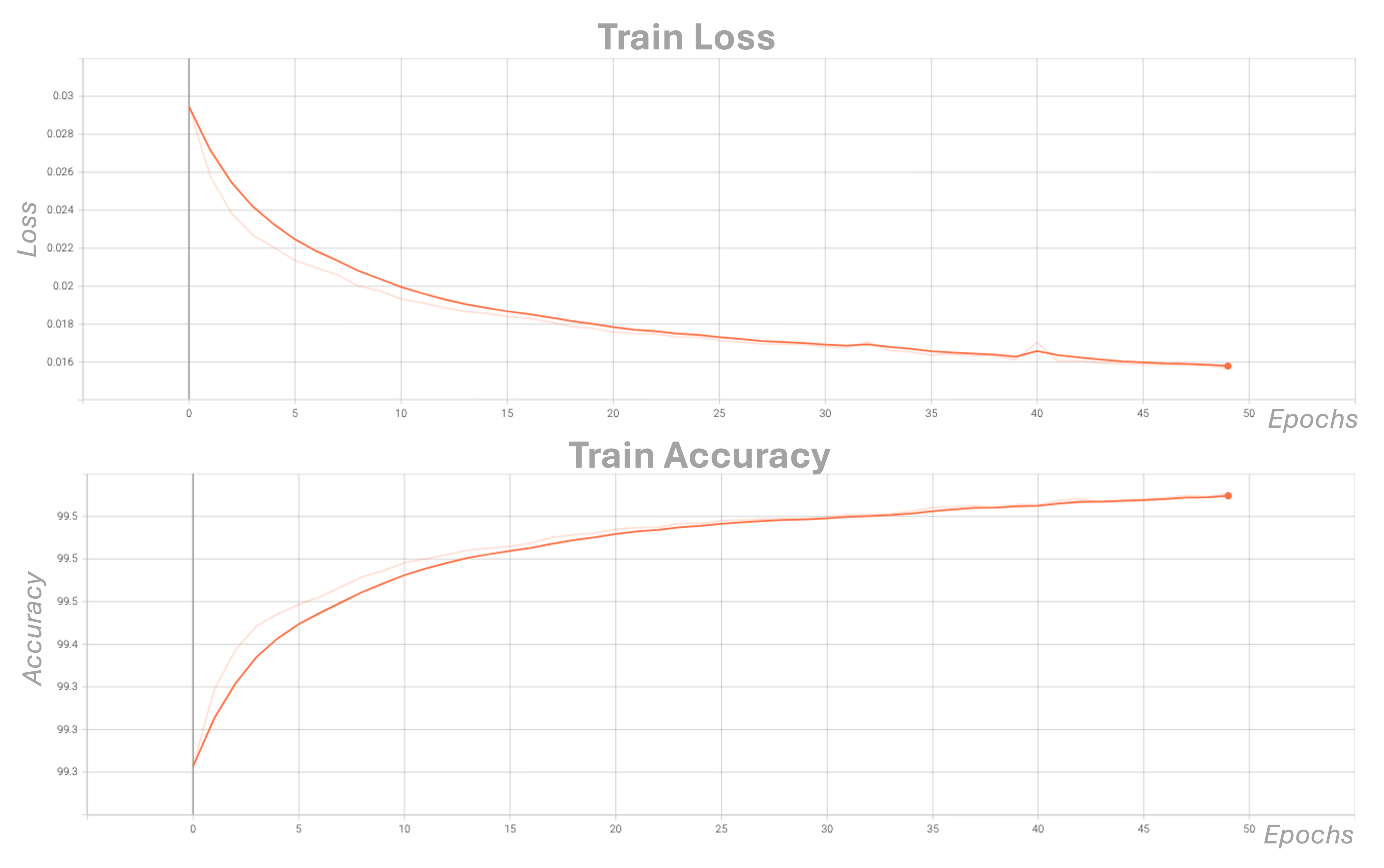}
    \caption{Evolution of model's loss (top), and accuracy score (bottom) during training.}
    \label{fig:train}
\end{figure}

Post training, the model was able to achieve an accuracy score of \textbf{99.58}\% in the train set, with the last epoch's loss being equal to \textbf{0.0157}, as seen in Figure \ref{fig:train}. 

\begin{figure}[!ht]
    \centering
    \includegraphics[width=0.95\columnwidth]{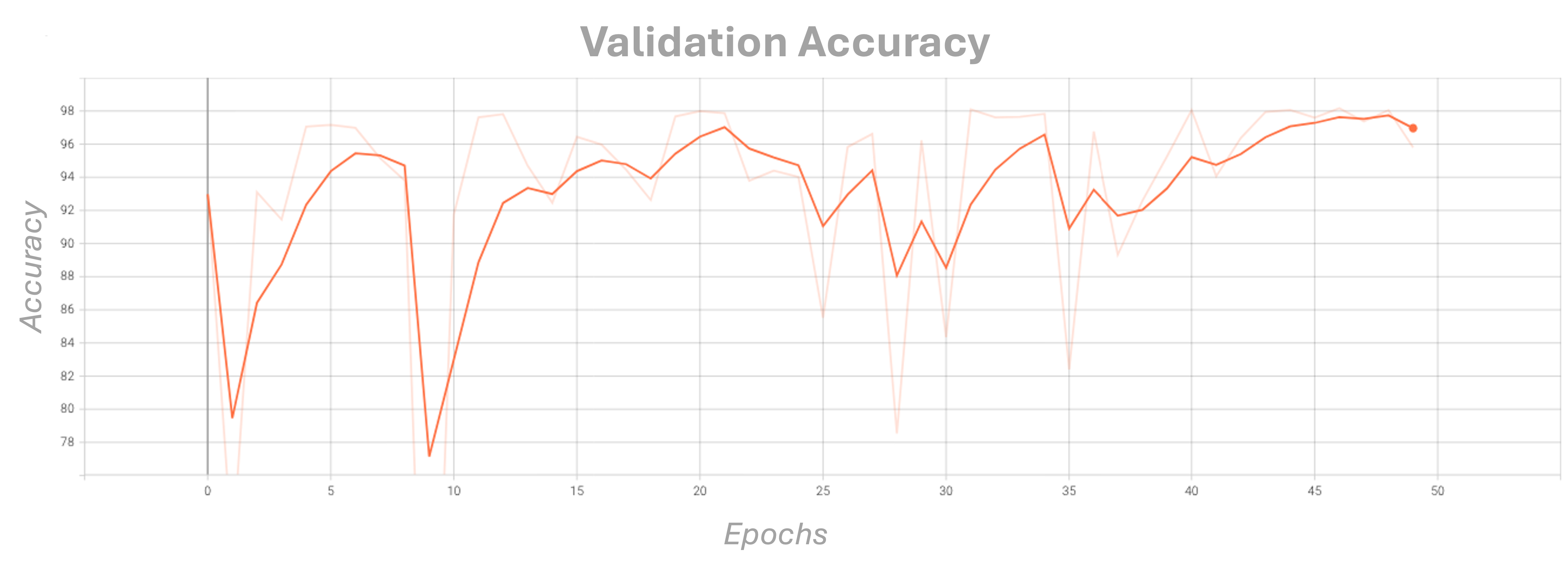}
    \caption{The model's performance in the validation set, during training, with the best accuracy score achieved in epoch 35.}
    \label{fig:validation}
\end{figure}

In Figure \ref{fig:validation}, one may notice that the model achieved peak performance in the validation set at epoch 35, with an accuracy score of \textbf{98.05}\%. 

\subsection{Post-Processing}

The proposed model is capable of processing \textit{330} lines of shape \(640\times224\) per second. In order to take advantage by the spatial domain and to acquire the final classification map, each of the predicted lines is accumulated in a 3D buffer until the whole set of lines post-processed together. 

\textbf{Median filter:} a kernel size 5, is applied to the reconstructed classification map. In this manner, misclassifications that reassemble the \textit{Salt \& Pepper} noise are correct by substituting them with the median value of the surrounding \(5\times5\) region, as described in Equation \ref{eq:median}. 

\begin{equation}
\label{eq:median}
    I'(x, y) = \text{median} \{ I(i, j) \mid (i, j) \in W(x, y) \}~~, 
\end{equation}
where \(I(x, y)\) is the original image, \(I'(x, y)\) is the filtered image, and\(W(x, y)\) is the neighborhood window centered at \((x, y)\).

\textbf{Morphological opening and closing:} the application of such filters enhances the overall quality of the segmentation by removing small noise artifacts and refining object boundaries. The opening filter effectively eliminates small, isolated regions of misclassified pixels, while the closing filter fills in small gaps and smoothens the contours of classified regions, \cite{serra1982image,de1984morphological}. Morphological filters presented respectively in Equations \ref{eq:open}, \ref{eq:close} 

\begin{equation}
\label{eq:open}
    I \circ B = (I \ominus B) \oplus B~~, 
\end{equation}
where \(I\) is the original image, \(B\) is the structuring element, \(\ominus\) denotes the erosion operation, \(\oplus\) denotes the dilation operation. 

\begin{equation}
\label{eq:close}
    I \bullet B = (I \oplus B) \ominus B~~,
\end{equation}
where \(I\) is the original image, 
\(B\) is the structuring element, 
\(\oplus\) denotes the dilation operation, 
\(\ominus\) denotes the erosion operation.

Together, these morphological operations improve the structural integrity of the classification map, leading to more accurate and visually coherent segmentation results. 

\section{Results \& Discussion}
\label{sec:results}
In this section the results of the proposed work on pixel-level material classification of hyperspectral images are presented, the structure of the sections is as follows a) overall performance, b) capacity to analyse randomly shredded objects, c) ability to distinguish mixed overlapping materials and d) limitations are discussed.

\subsection*{\textbf{Overall Performance}}
As discussed in Section \ref{sec:dataset} the test-set presented in Table \ref{tab:without_black_pixels} includes images of all the HDPE, PET, PP, PS classes, which however have not been utilised in the training subset. In this manner, guarantee is provided that no same object or pixel is simultaneously evident in both train and test data.

\begin{figure}[!ht]
    \centering
    \includegraphics[width=0.95\columnwidth]{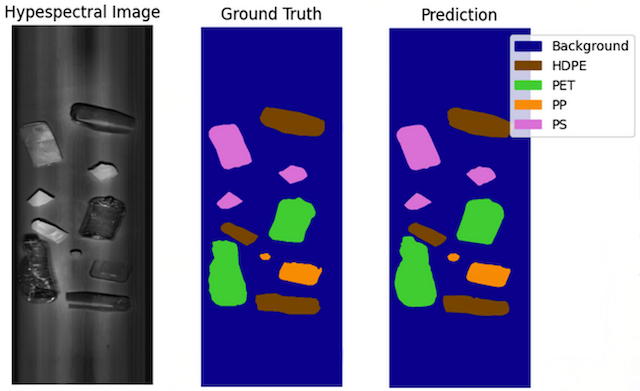}
    \caption{The false-colour version of the HS images, the ground truth mask, as well as the generated classification map from the proposed model.}
    \label{fig:hs-inference-shape}
\end{figure}

The hyperspectral images of this dataset are used by P1CH classifier in order to predict the material class of each pixel. Figure \ref{fig:hs-inference-shape} shows the original image as well as the ground truth and the prediction of each pixel. 

\begin{figure}[!ht]
    \centering
    \includegraphics[width=0.95\columnwidth]{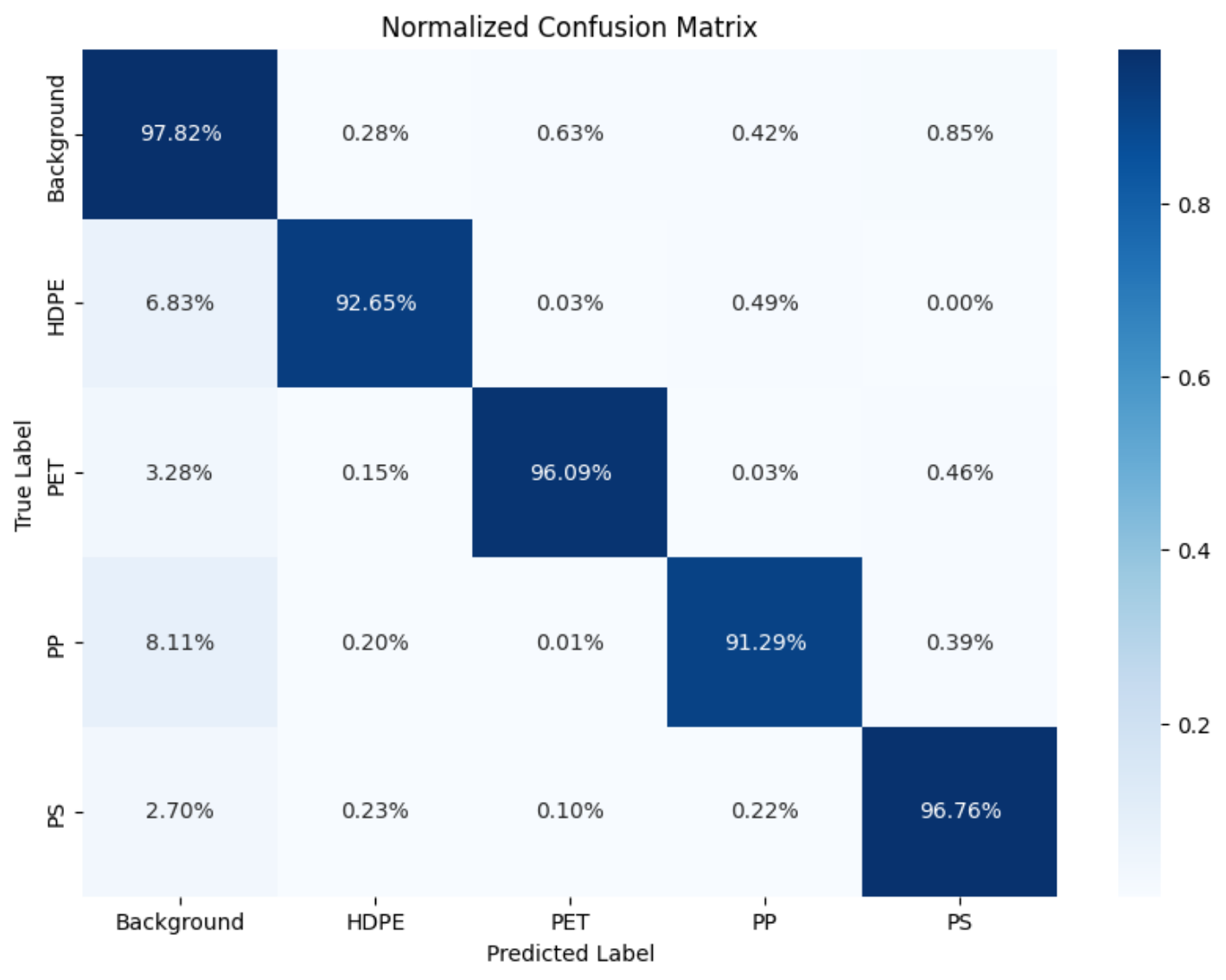}
    \caption{Confusion Matrix describing models performance in classifying materials on pixel-level.}
    \label{fig:confusionM}
\end{figure}

To quantify the model's performance, the accuracy score was calculated for the whole test set in pixel level thus providing a detailed evaluation on the model's classification capacity. The overall accuracy achieved by the model is \textbf{97.44}\%. The confusion matrix is presented in Figure \ref{fig:confusionM}, summarizing the classification performance among different material classes. The cell values in the confusion matrix are row-wise normalized, i.e. normalized with respect to the total number of samples in each class.
By observing the confusion matrix of Figure \ref{fig:confusionM} is easy to notice that less than \textit{1}\% of the total pixels of each class is falsely classified as another material and most of the errors are misclassifications between materials and the Background.

To further validate the performance of the proposed P1CH model, the state-of-the-art HybridSN methodology \cite{roy2019hybridsn,roy2020attention} was replicated and evaluated on the same test dataset. This approach ensures a direct and fair comparison under identical conditions. The overall accuracy achieved by the HybridSN model on this dataset is significantly lower, at \textbf{21.81}\%. In contrast, the P1CH model achieves an accuracy of \textbf{97.44}\%, along with superior recall and Kappa score. These results, summarized in Table~\ref{tab:comparison_metrics}, underline the robustness and reliability of the proposed P1CH model in pixel-level material classification tasks.

\begin{figure}[!ht]
    \centering
    \includegraphics[width=0.95\columnwidth  ]{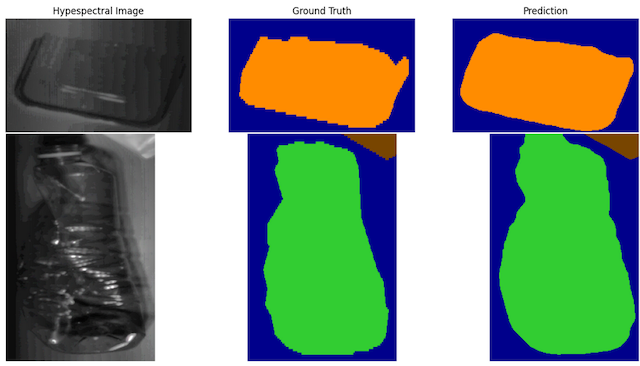}
    \caption{Two zoomed crops of Figure \ref{fig:hs-inference-shape}. In the first row the PP sample originally located on the bottom right-side of Figure \ref{fig:hs-inference-shape} is depicted with the GT and Predicted masks. In the second row the PET sample originally located on the left-side of Figure \ref{fig:hs-inference-shape} is presented.}
    \label{fig:zoom}
\end{figure}

\begin{table}[h!]
\centering
\caption{Performance Metrics Comparison between P1CH and HybridSN Models}
\label{tab:comparison_metrics}
\begin{tabular}{|l|c|c|}
\hline
\textbf{Metric}       & \textbf{P1CH}       & \textbf{HybridSN}     \\ \hline
Mean Accuracy (\%)         & 97.44              & 21.81                \\ \hline
Mean Recall (\%)                & 85.99              & 7.86                 \\ \hline
Mean Kappa Score           & 0.9295             & -0.0795              \\ \hline
Mean Inference Time (sec)  & 5.06               & 34.29                \\ \hline
\end{tabular}
\end{table}

The performance metrics summarized in Table~\ref{tab:comparison_metrics} further underscore the effectiveness of the proposed P1CH classifier. Notably, the accuracy achieved by P1CH (\textbf{97.44\%}) is significantly higher than that of the HybridSN model (\textbf{21.81\%}), highlighting its superior capacity to correctly classify pixels across the test set. In addition to accuracy, the recall value of P1CH (\textbf{85.99\%}) demonstrates its robustness in identifying material classes with high consistency, in stark contrast to the HybridSN model, which achieved a recall of only \textbf{7.86\%}. The Kappa score further illustrates the reliability of the P1CH model, with a value of \textbf{0.9295} compared to the negative score produced by HybridSN, reflecting its struggles with pixel-level classification tasks. 

Beyond classification accuracy, the computational efficiency of P1CH is a notable advantage. The proposed model achieves an inference speed of \textbf{200 lines per second (5.06 seconds per image)}, far surpassing HybridSN's \textbf{29 lines per second (34.29 seconds per image)}. This substantial difference not only demonstrates the lightweight design of P1CH but also makes it a practical choice for almost real-time applications, where rapid and reliable pixel-level classification is essential. With these results, P1CH emerges as a robust, accurate, and highly efficient solution for hyperspectral image pixel-wise classification in demanding environments.

\begin{figure*}[!ht]
    \centering
    \includegraphics[width=0.9\textwidth ,height=0.42\textheight]{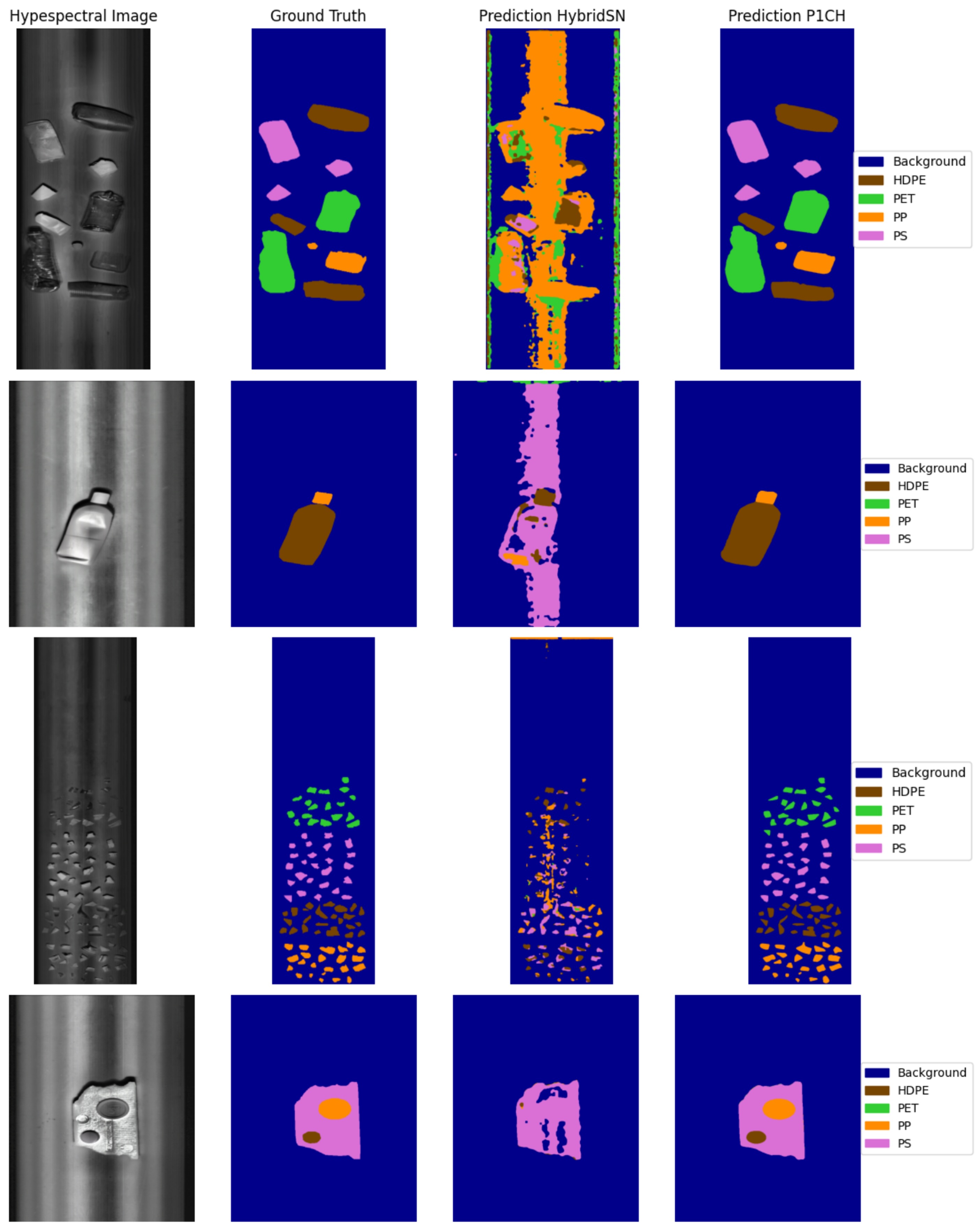}
    \caption{From left to right: The false-colour HS image, the Ground Truth mask, the classification maps generated by HybridSN model ,as well as the generated classification map, by the P1CH Classifier, in the aforementioned scenarios.}
    \label{fig:comparison_results}
\end{figure*}

A visual comparison of predictions by both models, on the test set, is presented in Figure~\ref{fig:comparison_results}. The HybridSN model struggles with pixel-level classification accuracy, particularly at object boundaries, and often misclassifies a large portion of pixels. In contrast, the P1CH model demonstrates precise segmentation and classification, even in challenging scenarios involving overlapping objects or small irregular shapes. These findings validate the effectiveness of the P1CH model and its potential applicability in real-world hyperspectral imaging tasks.

\subsection*{\textbf{Focusing on the borders}}
Moreover, it can be calculated that \textit{97.45}\% of the error comes in the borders between objects and the background, as suggested by the relatively low recall score. After more careful inspection is not easy to conclude that the error is actually on the predictions of P1CH and not on the ground truth masks. Figure \ref{fig:zoom} presents two zoomed-in crops of Figure \ref{fig:hs-inference-shape}, where it is clearly shown that in both cases the P1CH has the capacity to generate much smoother and precise masks for the respective object while simultaneously predicting correctly the material class. Comparing, also, the zoomed crops one can realize the mistake made on the top right side of the PP sample's mask, as well as at the PET bottle's spout. In both cases, P1CH was capable of predicting a more precise than the semi-automatically generated mask.

Therefore, given the aforementioned observations if the misclassifications on the borders are excluded from error calculations, the updated error rate is \textit{0.0653}\% and hence the total pixel-wise accuracy of P1CH classifier is \textbf{99.94}\%!

\subsection*{\textbf{Shredded materials}}

% \begin{figure*}[!ht]
%     \centering    \includegraphics[width=0.9\textwidth]{imgs/results_p2.png}
%     \caption{Model's output in challenging, cluttered scenes, where the objects are small with irregular shapes and similar textures.}
%     \label{fig:shredded}
% \end{figure*}

Beyond overall accuracy metrics, evaluating the model's effectiveness in classifying randomly shaped plastic samples is critical, as these shapes present significant challenges for RGB computer vision models and the same time is needed in various real-life applications. In this manner, an ensemble of plastic samples was collected and shredded in small, irregular pieces, and then placed on the conveyor belt. P1CH classifier demonstrates remarkable capabilities in classifying on pixel-level the different material classes, clearly proving its superior performance compared to traditional RGB-based instance segmentation algorithms.

Traditional RGB models often struggle with such high variability due to their reliance on surface-level features like colour, texture and shape, which are limited in the case of small shredded samples. On the contrary, the proposed hyperspectral imaging approach is trained to purely exploit the rich spectral content of each pixel in the HS image, hence successfully tackling the challenge of small and irregularly shaped objects. Another key advantage of the proposed model is its resilience to noise and artifacts commonly found in RGB images. While traditional models can be easily misled by variations in lighting and surface texture, the spectral information utilised by the proposed model provides a more stable basis for classification.

\begin{figure}[!ht]
    \centering    \includegraphics[width=1\columnwidth]{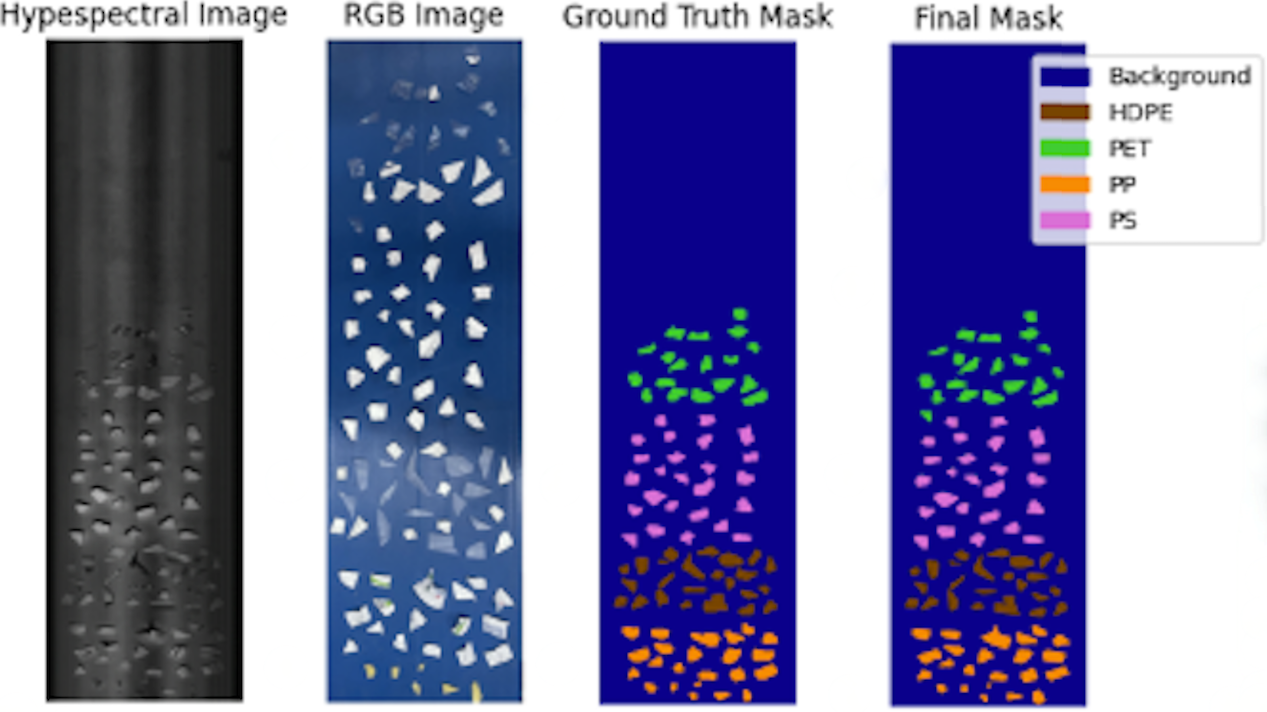}
    \caption {From left to right: The false-colour HS image, the RGB equivalent image, the Ground Truth mask, as well as the generated classification map, by the P1CH Classifier, in a challenging, cluttered scene, where the objects are small with irregular shapes and similar textures.}
    \label{fig:shredded}
\end{figure}

In Figure \ref{fig:shredded} the HS and the respective RGB images of the shredded plastic samples of HDPE, PET, PP and PS are depicted. Along with the aforementioned images, the Ground Truth and the Predicted masks are presented in the same figure. The achieved accuracy in this specific image is \textit{98.9}\%, with all the misclassifications falling under the borders case described in the previous section.  

Visual comparison between the ground truth and the predicted mask indicates a high level of agreement in the classification of the plastic fragments. Each class—HDPE, PET, PP, and PS—is distinctly identified and accurately located in the predicted mask. In detail, even though the samples have similar colours regardless of their class, being either white or transparent, the model was able to correctly classify all the samples to their respective category, as indicated in the ground truth mask. This result underscores the robustness of P1CH is \textbf{invariant to colour and to size} of the objects by solely relying on the spectral content of each individual pixel. 

\begin{figure}[!ht]
    \centering
    \includegraphics[width=0.95\columnwidth, height = 7cm ]{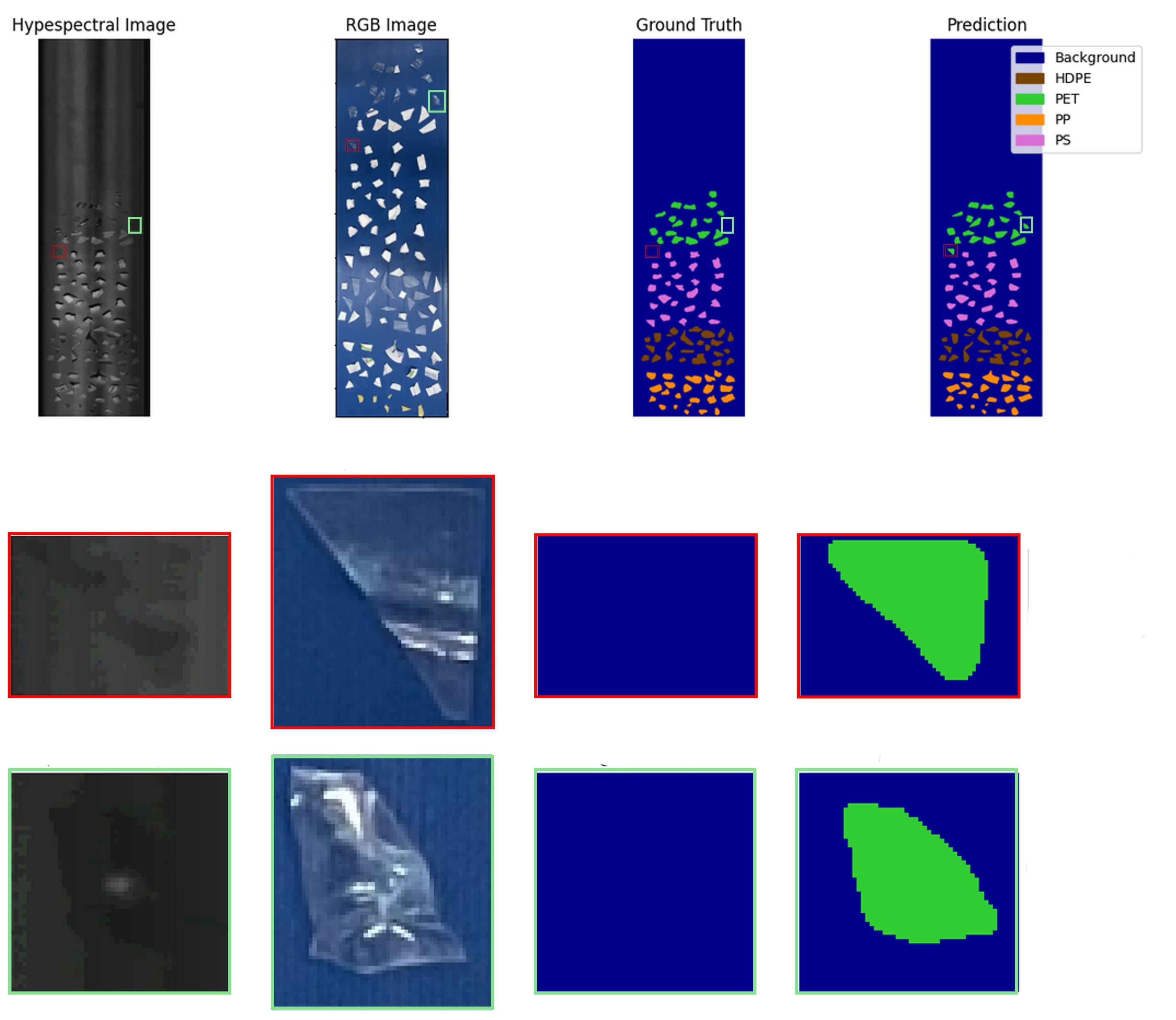}
    \caption{Zoomed-in crops of Figure \ref{fig:shredded}, highlighting two small PET shards that were mistakenly omitted from the labelling process (denoted with red and green boxes), and yet were detected by the model.}
    \label{fig:undetected}
\end{figure}

Even though the points just presented are important, the most impressive observation from this experiment is the model's ability to detect objects that were not even visually identified during the labeling process of the HS images. Figure \ref{fig:undetected} specifies two regions in the HS image, denoted with red and green boxes respectively, in which there are two small shards of PET samples. Inspecting the zoomed-in crops presenting in Figure \ref{fig:undetected}. It is almost impossible to detect those objects, yet the proposed HS approach clearly identified these two objects as PET. This outcome is only possible through the analytical processing of the spectral information of pixels in those regions, resulting in very precise masks and ultimately correctly classified samples. 
\begin{figure*}[!ht]
    \centering
    \includegraphics[width=0.9\textwidth]{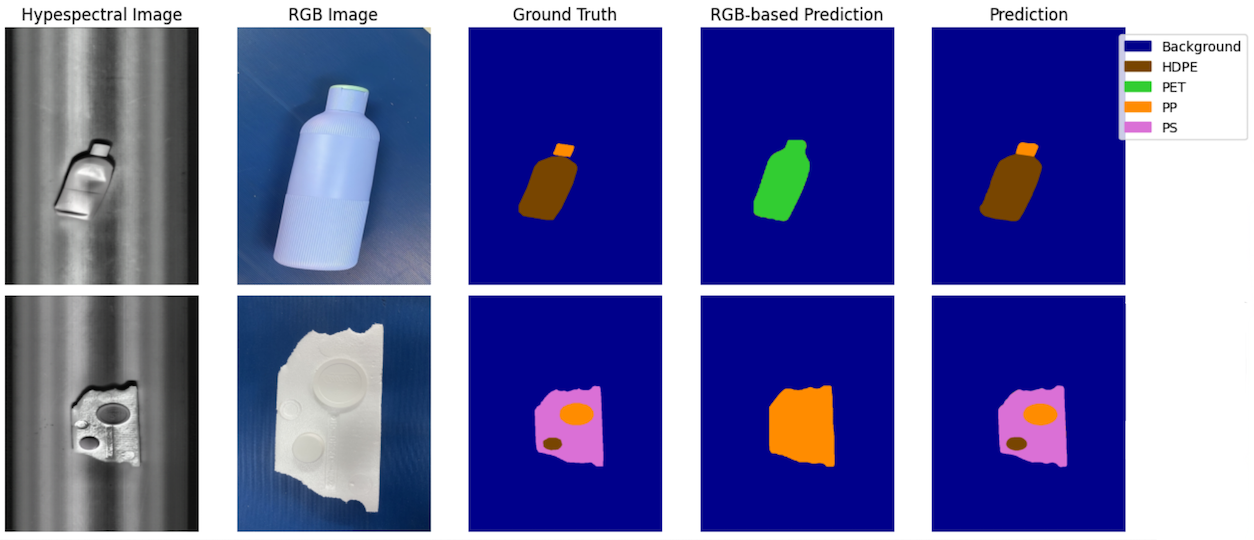}
    \caption{From left to right: The false-colour HS image, the RGB equivalent image, the Ground Truth mask, the classification maps generated by an RGB-based model using the RGB equivalent images,as well as the generated classification map, by the P1CH Classifier, in the scenario of mixed or overlapping materials.}
    \label{fig:cluttered}
\end{figure*}

% \begin{figure*}[!ht]
%     \centering
%     \includegraphics[width=0.9\textwidth]{imgs2/results_p1.png}
%     \caption{From left to right: The false-colour HS image, the RGB equivalent image, the Ground Truth mask, the classification maps generated by an RGB-based model using the RGB equivalent images,as well as the generated classification map, by the P1CH Classifier, in the scenario of mixed or overlapping materials.}
%     \label{fig:cluttered}
% \end{figure*}

\subsection*{\textbf{Mixed overlapping materials}}

In addition to the challenge posed by small irregularly shaped objects, conventional RGB-based detection models often fail to accurately identify distinct samples that are either attached to one another or overlapping. In this subsection, an analysis is conducted to evaluate the model's capacity to precisely segment samples of different material class, while accurately classifying at the same time each individual pixel to its respective class. 
To this end, two experiments were executed, and in Figure \ref{fig:cluttered} the HS and RGB images and the respective masks of the utilised objects are depicted. To further highlight the proposed model's ability to accurately classify materials and segment their instances within an image, a fine-tuned on the specific classes RGB-based instance segmentation model was also employed to segment the RGB images. The predictions of the RGB-based model are depicted in the 4th column of Figure \ref{fig:cluttered}.

In the first experiment, a common, commercially available, light blue shampoo bottle was selected. From a visual point of view, as seen in the RGB image, the body of the shampoo container and its lid appear almost identical making it impossible, even for a human, to realize that those two parts are different types of plastic, i.e. HDPE and PP. From the first row in Figure \ref{fig:cluttered}, one can easily observe that the RGB-based model was able to identify the contour of greater complex of objects, but not the individual parts, i.e. body and lid. Moreover, the RGB-based model failed to predict the class of both samples, assigning them to the PET class. Finally, due to shadows and non uniform illumination of the scene, the RGB generated mask lacks in precision as it includes pixels of the background too. On the contrary, the proposed HS approach, not only generated highly precise masks for both the body and the lid, but also is capable of classifying those two parts in their respective class. The overall accuracy of the prediction for this specific test case is \textit{98.8}\%, with disagreements in a pixel's class being evident only between background and not the material classes. 

In the second experiment, the case of overlapping materials with very similar (white) colour was examined. To this end, a white PS flat surface was selected, on top which a white PP (top right) and a white HDPE (bottom left) lid were placed. The respective masks and images are depicted in the second row of Figure \ref{fig:cluttered}. Looking at the RGB image, no distinctive textures can be detect for the two lids, while some minor changes in texture may be identified between the lids and the PS surface. Therefore, given the uniformity in colour and the very low variance in texture, as anticipated, the RGB-based model fail to detect all three objects apparent in the scene. Once again, the generated mask contains the greater complex of samples with no regard to the two lids. Moreover, the class prediction of the RGB-based is inaccurate since it considers the PS surface as PP. In contrast to the RGB case, the proposed HS model effectively utilises the rich spectral signature encoded in each pixel being able to precisely segment all three components of the image, while classyfing materials on pixel-level with \textit{99.54}\% accuarcy.

The results presented above underscore a pivotal advancement in material classification and segmentation capabilities. The most remarkable achievement demonstrated here is the proposed HS model's ability to accurately identify and classify distinct, overlapping samples with complex boundaries. This marks a significant breakthrough, as conventional RGB-based models consistently fail under these conditions, misidentifying materials and producing imprecise masks. The superior performance of P1CH classifier achieving up to \textbf{99.54}\% accuracy even in the presence of overlapping objects, illustrates a transformative improvement in hyperspectral imaging applications, paving the way for more sophisticated and reliable material detection and sorting systems in real-world environments.

\subsection*{\textbf{Limitations}}

\begin{figure*}[!ht]
    \centering
    \includegraphics[width=0.9\textwidth]{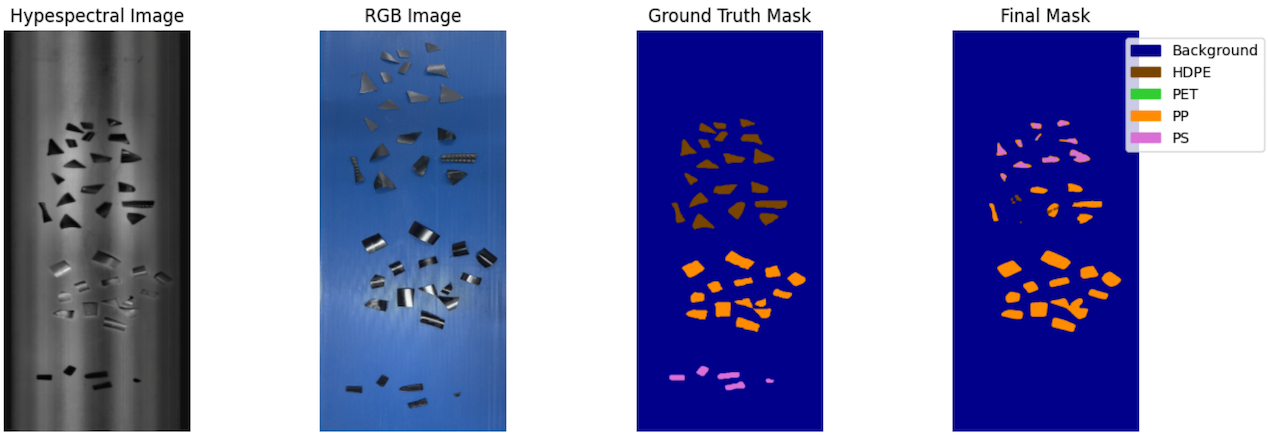}
    \caption{From left to right: The false-colour HS image, the RGB equivalent image, the Ground Truth mask, as well as the generated classification map, by the P1CH Classifier, in the case of black plastics.}
    \label{fig:black}
\end{figure*}

Despite the impressive results acquired in the previous experiments, this work intendeds to also underline the limitations of the HS imaging in classifying materials. To this end, the last experiment involves the analysis of dark-coloured, irregularly shaped samples. For the needs of this experiment, black and dark-coloured HDPE, PP and PS objects were cut in random small fractions and placed on top of the conveyor belt. The respective images and masks are presented in Figure \ref{fig:black}. This subsection delves into the acquired results of the analysis of dark samples, discussing also the reasons why the model performs poorly in that case.

Examining the predictions, in the last column of Figure \ref{fig:black} it is concluded that the model performs very poorly in the case of black plastics. In detail, only PP samples were are correctly classified, although lacking in mask precision. The PS samples were not completely undetected, while some HDPE samples were identified by the model but once again misclassified as PS or PP. The performance of the proposed model in terms of accuracy without taking into consideration the background is  equal to merely \textit{39.69}\%.

\begin{figure}[!ht]
    \centering
    \includegraphics[width=0.95\columnwidth]{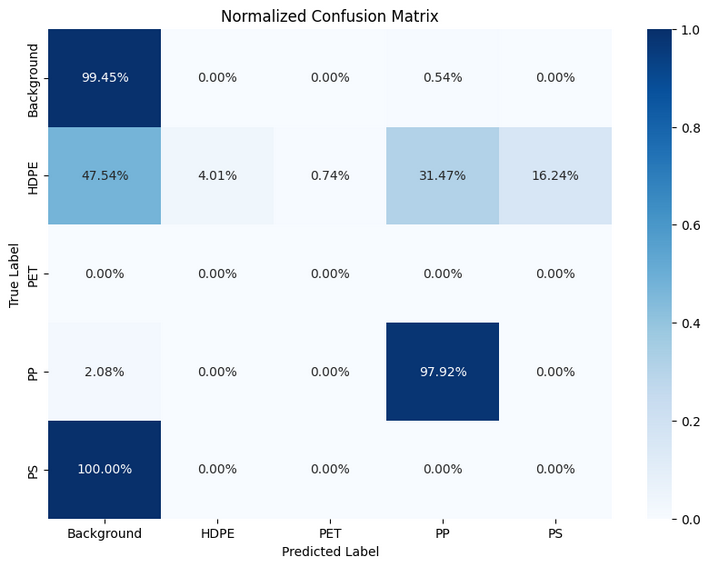}
    \caption{Normalized Confusion Matrix for the predictions of the model in the case of black and dark-coloured samples.}
    \label{fig:cfmBlack}
\end{figure}

The confusion matrix presented in Figure \ref{fig:cfmBlack} specifically summarizes the models performance in the case of black samples. In detail, it is confirmed that all the PS pixels were classified as background, i.e. they were not detected by the model at all. Moreover, from this confusion matrix one can see that indeed the PP samples were accurately classified, with the models precision in that case dropping to \textit{75}\% since HDPE samples are also misclassifies as PP. Finally, for the HDPE case, only \textit{4.01}\% of the total samples were correctly classified, with the model's recall for the specific case being equal to \textit{46}\%.

\begin{figure}[!ht]
    \centering
    \includegraphics[width=0.95\columnwidth]{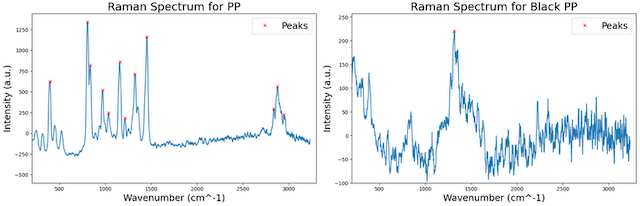}
    \caption{A comparison between the spectra acquired from a black (left) and a white (right) PS sample. A significant drop in the signal's amplitude can be observed in the case of black PS. Also one can notice that most of the indicative peaks for PS are non-existent in the case of black PS.  }
    \label{fig:blackSpectrum}
\end{figure}

At this point, it of paramount importance to explain the reasons that lead in this drop in the model's performance. Specifically, this phenomenon should mainly be attributed to the nature of dark-coloured materials, rather than being considered a model's deficiency. In detail, the observed dark or even black colour of an object is the result of complete or almost complete absorption of the incident radiation. In this manner, the reflected radiation, which is captured by the HS cameras' sensor, is of very low intensity; hence resulting in a very weak digital signal. Therefore, given the low amplitude of the captured signal, the PSNR is consequently also low, and so is the variability of the spectrum. All the above, result in very similar -almost identical-, noisy features of the model's  input vector, thus rendering the model incapable of properly analyzing each pixel's spectrum and ultimately correctly classifying them in their respective classes. An example of the spectrum of a black and a white plastic of the same material are presented in Figure \ref{fig:blackSpectrum}. Can be easily observed that the spectrum of the black-colored object is much more noisy with $10$ times lower intensity.

Although, it should be highlighted that black plastics of any material class were totally absent from the training set. The exploration of techniques that would probably lead to mitigate the pure performance of computer vision with HS imaging on black objects is beyond the scope of this paper, but opens and intriguing research direction on the intersection between material science and artificial intelligence.

% An additional significant finding from this study, extending beyond the specific test set images, is the reduced dependency on conventional spectral calibration techniques. In this work the utilization of the spectral calibration matrix was omitted and yet, as mentioned in \textit{Overall Performance} subsection, outstanding results were achieved with an overall accuracy score of \textbf{99.94}\% . Moreover, in this work the conventional black-white reference normalization was substituted with a simpler and more cost effective methodology. 

\section{Conclusion \& Future Work}
\label{sec:conclusion}
In this paper a lightweight \textit{1D Convolutional  Hyperspectral Classification}, system, P1CH, is proposed. The developed model generates highly detailed and precise classification maps. Unlike conventional RGB  methods, the proposed algorithm utilises the spectral information encoded in hyperspectral image's pixel, allowing to detect and correctly classify objects in challenging scenarios.

For the training and validation purposes two sets of hyperspectral images were generated. Both of the dataset splits contain images of \textit{HDPE, PET, PP} and \textit{PS} samples. Each of the images in the training set, contain explicitly objects of one class, while the images in the test set aim to reproduce challenging conditions, hence depicting very small irregularly shaped objects, overlapping objects, and mixed materials with similar to identical texture. A simplified, cost-efficient spectral calibration and normalization technique is, also, proposed in this work that do not requires specialized hardware. The model was validated on the aforementioned test set and it achieved an overall accuracy of \textbf{99.54}\%. Moreover, to explore the limitations of the P1CH Classifier the scenario of classifying black samples was explored, where it was noticed that the model struggled to properly classify the samples. 

On the one hand, this work presents the capabilities of computer vision with Hyperspectral imaging. On the other hand raised limitations and problems that are absolutely worth systematic research effort. Further investigation on the problem of black or dark-coloured samples. Study the capabilities on more extensive spectral range. Moreover, the model's robustness in various illumination conditions is set to be further examined. Furthermore, dimensionality reduction and compression problems have to be examined, as well as the effects such operations might have on the learning performance. Finally, the utilisation of the spatial information encoded in hyperspectral images, shall also be examined in future studies.

% or




%\bibliographystyle{IEEEtran}
% \section*{Acknowledgements}

% This research was financially supported by the European Union’s Horizon Europe research and innovation program under grant agreement No 101138789 (project W2W).
\bibliographystyle{IEEEtran}
\bibliography{references}

\vskip8pt

\end{document}